\documentclass[fleqn]{llncs}
\usepackage{igpc}

\title{An Interface Group for Process Components%
       \thanks{This research was partly carried out in the framework of
               the  Jacquard-project Symbiosis, which is funded by the
               Netherlands Organisation for Scientific Research (NWO).}}
\author{J.A. Bergstra \and C.A. Middelburg}
\institute{Programming Research Group, University of Amsterdam, \\
           P.O.~Box~41882, 1009~DB~Amsterdam, the Netherlands \\
           \email{J.A.Bergstra@uva.nl,C.A.Middelburg@uva.nl}}

\begin{document}

\maketitle

\begin{abstract}
We take a process component as a pair of an interface and a behaviour.
We study the composition of interacting process components in the
setting of process algebra.
We formalize the interfaces of interacting process components by means
of an interface group.
An interesting feature of the interface group is that it allows for
distinguishing between expectations and promises in interfaces of
process components.
This distinction comes into play in case components with both client and
server behaviour are involved.
\begin{classcode}
D.2.1, D.2.2, D.2.4, F.1.2, F.3.1.
% MSC 2000: 68Q85, 68N30, 20K99
\end{classcode}
\end{abstract}

\section{Introduction}
\label{sect-introduction}

Component interfaces are a practical tool for the development of all but
the most elementary architectural designs.
In~\cite{BP07a}, interface groups have been proposed as a means to
formalize the interfaces of the components of financial transfer
architectures.
The interface groups introduced in that paper concern component
behaviours of a special kind, namely financial transfer behaviours of
units of an organization.
In this paper, we introduce an interface group which concerns behaviours
of a more general kind, namely behaviours that can be viewed as
processes specifiable in the process algebra known as
\ACP~\cite{BW90,Fok00}.

An interface group is a commutative group intended for describing and
analysing interfaces.
The interface group introduced in this paper concerns interfaces of
process components that request other components to carry out methods
and grant requests of other components to carry out methods.
The interfaces in question represent the abilities to grant requests
that are expected from other components and the abilities to make
requests that are promised to other components.
The ability to make a certain request and the ability to grant that
request are considered to cancel out in interfaces.
Thus, having an empty interface is a sufficient condition on a process
component for being a closed system.
Interfaces as modelled by the interface group introduced in this paper
have less structure than the signatures used as interfaces in module
algebra~\cite{BHK90a}.
However, module algebra does not allow for distinguishing between
expectations and promises in interfaces of components.
In point of fact, it has a bias towards composing components whose
interfaces concern promises only.

We also present a theory about process components of which the interface
group introduced forms part.
Like any notion of component, the notion of process component underlying
this theory combines interface with content: a process component is
considered a pair of an interface and a behaviour.
Processes as considered in \ACP\ are taken as the behaviours of process
components.
Therefore, the theory concerned is a development on top of \ACP.
However, additional assumptions are made about the actions of which the
processes are made up.
Three kinds of actions are distinguished: the acts of making requests
referred to above, the acts of granting requests referred to above, and
the acts of carrying out methods which result from making a request and
granting that request at the same time.
The use of the presented theory about process components is illustrated
by means of examples.
A model of the theory is constructed, using a notion of bisimilarity for
process components.

In the presented theory about process components, composition of process
components is in general not associative.
Little can be done about this because turning a process into a component
by adding an interface to it inevitably results in encapsulation of the
process.
However, composition of process components is associative when a certain
condition on the process components in question is fulfilled.
We couch this in a special associativity axiom for component
composition.

In the presented theory about process components, processes reside at
places, called loci, and requests and grants are addressed to the
processes residing at a certain locus.
If the processes that are taken as the behaviours of process components
are looked at in isolation, it may be convenient to abstract from the
loci at which they reside.
This abstraction gives rise to another kind of processes.
We treat this kind of processes, referred to as localized processes, as
well.

A system composed of a collection of process components is a closed
system if the actions that make up its behaviour include neither acts of
making requests nor acts of granting requests.
It is generally undecidable whether a system composed of a collection of
process components is a closed system.
This state of affairs forms part of the motivation for developing the
theory about process components presented in this paper.
In the presented theory, having an empty interface is a sufficient
condition for being a closed system and it is decidable whether an
interface is empty.

The structure of this paper is as follows.
First, we review \ACP\ (Section~\ref{sect-ACP}) and guarded recursion in
the setting of \ACP\ (Section~\ref{sect-REC}), and present the actions
that make up the processes being considered in later sections
(Section~\ref{sect-ACPcc}).
Next, we introduce a theory about integers (Section~\ref{sect-INT}) and
a theory about interfaces (Section~\ref{sect-IFGcc}).
Then, we extend \ACP, using the theories just introduced, to a theory
about process components (Section~\ref{sect-ACC}).
Following this, we go into the matter that component composition is in
general not associative (Section~\ref{sect-non-assoc}) and discuss the
connection between empty interfaces and closed systems
(Section~\ref{sect-closed-systems}).
After that, we give two examples of the use of the presented theory
about process components
(Sections~\ref{sect-example} and~\ref{sect-another-example}).
Thereupon, we introduce a notion of bisimilarity for process components
(Section~\ref{sect-bisim}) and construct a model of the presented theory
about process components using this notion of bisimilarity
(Section~\ref{sect-model}).
Following that, we extend the theory about process components developed
so far with localized processes (Section~\ref{sect-localized}).
Finally, we make some concluding remarks
(Section~\ref{sect-conclusions}).

\section{Algebra of Communicating Processes}
\label{sect-ACP}

In this section, we shortly review \ACP\ (Algebra of Communicating
Processes), the algebraic theory about processes that was first
presented in~\cite{BK84b}.
For a comprehensive overview of \ACP, the reader is referred
to~\cite{Fok00}.
Although \ACP\ is one-sorted, we make this sort explicit.
The reason for this is that we will extend \ACP\ to a theory with four
sorts in Section~\ref{sect-ACC}.

In \ACP, it is assumed that a fixed but arbitrary finite set of
\emph{actions} $\Act$, with $\dead \not\in \Act$, has been given.
We write $\Actd$ for $\Act \union \set{\dead}$.
It is further assumed that a fixed but arbitrary commutative and
associative \emph{communication} function
$\funct{\commm}{\Actd \x \Actd}{\Actd}$, with $\dead \commm a = \dead$
for all $a \in \Actd$, has been given.
The function $\commm$ is regarded to give the result of synchronously
performing any two actions for which this is possible, and to be $\dead$
otherwise.

\ACP\ has one sort: the sort $\Proc$ of \emph{processes}.
To build terms of sort $\Proc$, \ACP\ has the following constants and
operators:
\begin{itemize}
\item
the \emph{deadlock} constant $\const{\dead}{\Proc}$;
\item
for each $a \in \Act$, the \emph{action} constant $\const{a}{\Proc}$;
\item
the binary \emph{alternative composition} operator
$\funct{\altc}{\Proc \x \Proc}{\Proc}$;
\item
the binary \emph{sequential composition} operator
$\funct{\seqc}{\Proc \x \Proc}{\Proc}$;
\item
the binary \emph{parallel composition} operator
$\funct{\parc}{\Proc \x \Proc}{\Proc}$;
\item
the binary \emph{left merge} operator
$\funct{\leftm}{\Proc \x \Proc}{\Proc}$;
\item
the binary \emph{communication merge} operator
$\funct{\commm}{\Proc \x \Proc}{\Proc}$;
\item
for each $H \subseteq \Act$,
the unary \emph{encapsulation} operator
$\funct{\encap{H}}{\Proc}{\Proc}$.
\end{itemize}
Terms of sorts $\Proc$ are built as usual for a one-sorted signature
(see e.g.~\cite{Wir90a,ST99a})
Throughout the paper, we assume that there are infinitely many variables
of sort $\Proc$, including $x$, $y$, $z$, $x'$, $y'$ and $z'$.

We use infix notation for the binary operators.
The following precedence conventions are used to reduce the need for
parentheses.
The operator ${} \altc {}$ binds weaker than all other binary operators
to build terms of sort $\Proc$ and the operator ${} \seqc {}$ binds
stronger than all other binary operators to build terms of sort $\Proc$.

Let $P$ and $Q$ be closed terms of sort $\Proc$, $a \in \Act$, and
$H \subseteq \Act$.
Intuitively, the constants and operators to build terms of sort $\Proc$
can be explained as follows:
\begin{itemize}
\item
$\dead$ can neither perform an action nor terminate successfully;
\item
$a$ first performs action $a$ and then terminates successfully;
\item
$P \altc Q$ behaves either as $P$ or as $Q$, but not both;
\item
$P \seqc Q$ first behaves as $P$ and on successful termination of $P$ it
next behaves as $Q$;
\item
$P \parc Q$ behaves as the process that proceeds with $P$ and $Q$ in
parallel;
\item
$P \leftm Q$ behaves the same as $P \parc Q$, except that it starts
with performing an action of $P$;
\item
$P \commm Q$ behaves the same as $P \parc Q$, except that it starts
with performing an action of $P$ and an action of $Q$ synchronously;
\item
$\encap{H}(P)$ behaves the same as $P$, except that actions from $H$ are
blocked.
\end{itemize}

We write $\Altc{i \in \cI} P_i$, where $\cI = \set{i_1,\ldots,i_n}$ and
$P_{i_1},\ldots,P_{i_n}$ are terms of sort $\Proc$, for
$P_{i_1} \altc \ldots \altc P_{i_n}$.
The convention is that $\Altc{i \in \cI} P_i$ stands for $\dead$ if
$\cI = \emptyset$.

The axioms of \ACP\ are the axioms given in Table~\ref{axioms-ACP}.%
\begin{table}[!t]
\caption{Axioms of \ACP}
\label{axioms-ACP}
\begin{eqntbl}
\begin{axcol}
x \altc y = y \altc x                                  & \axiom{A1} \\
(x \altc y) \altc z = x \altc (y \altc z)              & \axiom{A2} \\
x \altc x = x                                          & \axiom{A3} \\
(x \altc y) \seqc z = x \seqc z \altc y \seqc z        & \axiom{A4} \\
(x \seqc y) \seqc z = x \seqc (y \seqc z)              & \axiom{A5} \\
x \altc \dead = x                                      & \axiom{A6} \\
\dead \seqc x = \dead                                  & \axiom{A7} \\
{}                                                                  \\
{}                                                                  \\
\encap{H}(a) = a               \hfill \mif a \not\in H & \axiom{D1} \\
\encap{H}(a) = \dead               \hfill \mif a \in H & \axiom{D2} \\
\encap{H}(x \altc y) = \encap{H}(x) \altc \encap{H}(y) & \axiom{D3} \\
\encap{H}(x \seqc y) = \encap{H}(x) \seqc \encap{H}(y) & \axiom{D4}
\end{axcol}
\qquad
\begin{axcol}
x \parc y =
        (x \leftm y \altc y \leftm x) \altc x \commm y & \axiom{CM1} \\
a \leftm x = a \seqc x                                 & \axiom{CM2} \\
a \seqc x \leftm y = a \seqc (x \parc y)               & \axiom{CM3} \\
(x \altc y) \leftm z = x \leftm z \altc y \leftm z     & \axiom{CM4} \\
a \seqc x \commm b = (a \commm b) \seqc x              & \axiom{CM5} \\
a \commm b \seqc x = (a \commm b) \seqc x              & \axiom{CM6} \\
a \seqc x \commm b \seqc y =
                        (a \commm b) \seqc (x \parc y) & \axiom{CM7} \\
(x \altc y) \commm z = x \commm z \altc y \commm z     & \axiom{CM8} \\
x \commm (y \altc z) = x \commm y \altc x \commm z     & \axiom{CM9} \\
{}                                                                   \\
a \commm b = b \commm a                                & \axiom{C1}  \\
(a \commm b) \commm c = a \commm (b \commm c)          & \axiom{C2}  \\
\dead \commm a = \dead                                 & \axiom{C3}
\end{axcol}
\end{eqntbl}
\end{table}
CM2--CM3, CM5--CM7, C1--C3 and D1--D4 are actually axiom schemas in
which $a$, $b$ and $c$ stand for arbitrary constants of sort $\Proc$
(keep in mind that also the deadlock constant belongs to the constants
of sort $\Proc$) and $H$ stands for an arbitrary subset of $\Act$.

For the main models of \ACP, the reader is referred to~\cite{BW90}.

\section{Guarded Recursion}
\label{sect-REC}

In this section, we shortly review guarded recursion in the setting of
\ACP.

Not all processes in a model of \ACP\ have to be interpretations of
closed terms of sort $\Proc$.
Those processes may be definable over \ACP.
A process in some model of \ACP\ is \emph{definable} over \ACP\ if there
exists a guarded recursive specification over \ACP\ of which that
process is the unique solution.

A \emph{recursive specification} over \ACP\ is a set of recursion
equations $\set{X = t_X \where X \in V}$ where $V$ is a set of variables
of sort $\Proc$ and each $t_X$ is a term of sort $\Proc$ from the
language of \ACP\ that only contains variables from $V$.
Let $E$ be a recursive specification over \ACP.
Then we write $\vars(E)$ for the set of all variables that occur on the
left-hand side of an equation in $E$.
A \emph{solution} of a recursive specification $E$ is a set of processes
(in some model of \ACP) $\set{p_X \where X \in \vars(E)}$ such that
the equations of $E$ hold if, for all $X \in \vars(E)$, $X$ stands for
$p_X$.

Let $t$ be a term of sort $\Proc$ from the language of \ACP\
containing a variable $X$.
Then an occurrence of $X$ in $t$ is \emph{guarded} if $t$ has a subterm
of the form $a \seqc t'$ where $a \in \Act$ and $t'$ is a term
containing this occurrence of $X$.
Let $E$ be a recursive specification over \ACP.
Then $E$ is a \emph{guarded recursive specification} if, in each
equation $X = t_X \in E$, all occurrences of variables in $t_X$ are
guarded or $t_X$ can be rewritten to such a term using the axioms of
\ACP\ in either direction and/or the equations in $E$ except the
equation $X = t_X$ from left to right.
We are only interested in models of \ACP\ in which guarded recursive
specifications have unique solutions.

For each guarded recursive specification $E$ and each variable
$X \in \vars(E)$, we introduce a constant of sort $\Proc$ standing for
the unique solution of $E$ for $X$.
This constant is denoted by $\rec{X}{E}$.
We often write $X$ for $\rec{X}{E}$ if $E$ is clear from the context.
In such cases, it should also be clear from the context that we use
$X$ as a constant.

The additional axioms for recursion are given
in Table~\ref{axioms-ACP-REC}.%
\begin{table}[!t]
\caption{Axioms for recursion}
\label{axioms-ACP-REC}
\begin{eqntbl}
\begin{caxcol}
\rec{X}{E} = \rec{t_X}{E} & \mif X = t_X \in E
& \axiom{RDP}
\\
E \Implies X = \rec{X}{E} & \mif X \in \vars(E)
& \axiom{RSP}
\end{caxcol}
\end{eqntbl}
\end{table}
In this table, we write $\rec{t_X}{E}$ for $t_X$ with, for all
$Y \in \vars(E)$, all occurrences of $Y$ in $t_X$ replaced by
$\rec{Y}{E}$.
Both RDP and RSP are axiom schemas.
Side conditions are added to restrict the variables, terms and guarded
recursive specifications for which $X$, $t_X$ and $E$ stand.
The equations $\rec{X}{E} = \rec{t_X}{E}$ for a fixed $E$ express that
the constants $\rec{X}{E}$ make up a solution of $E$.
The conditional equations $E \Implies X = \rec{X}{E}$ express that this
solution is the only one.
RDP and RSP were first formulated in~\cite{BK86c}.

We write \ACP+\REC\ for \ACP\ extended with the constants standing for
the unique solutions of guarded recursive specifications and the axioms
RDP and RSP.

\section{\ACP\ for Cooperating Components}
\label{sect-ACPcc}

In this paper, we consider process components that cooperate by making
and granting requests to carry out methods.
The processes that are taken as the behaviours of these components are
not made up of arbitrary actions.
In this section, we introduce the instance of \ACP\ that is restricted
to the intended actions.
This instance is called \ACPcc\ (\ACP\ for Cooperating Components).

Three kinds of actions are distinguished in \ACPcc: active actions,
passive actions and neutral actions.
The active actions may be viewed as requests to carry out some method
and the passive actions may be viewed as grants of requests to carry out
some method.
Making a request to carry out some method and granting that request at
the same time results in carrying out the method concerned.
The initiative in carrying out the method is considered to be taken by
the process making the request.
This explains why the request is called an active action and its grant
is called a passive action.
The neutral actions may be viewed as the results of making a request to
carry out some method and granting that request at the same time.
A process that can perform active actions only behaves as a client and a
process that can perform passive actions only behaves as a server.

In \ACPcc, it is assumed that a fixed but arbitrary finite set $\Loci$
of \emph{loci} and a fixed but arbitrary finite set $\Meth$ of
\emph{methods} have been given.
A locus is a place at which processes reside.
Intuitively, a process resides at a locus if it is capable of performing
actions in that locus.
The same process may reside at different loci at once.
Moreover, different processes may reside at the same locus at once.
Therefore, we do not necessarily refer to a unique process if we refer
to a process residing at a given locus.

In \ACPcc, the set of actions $\Act$ consists of:
\begin{itemize}
\item
for each $f,g \in \Loci$ and $m \in \Meth$, the \emph{active action}
$f.m \at g$;
\item
for each $f,g \in \Loci$ and $m \in \Meth$, the \emph{passive action}
$\passive f.m \at g$;
\item
for each $f,g \in \Loci$ and $m \in \Meth$, the \emph{neutral action}
$f.m \uat g$.
\end{itemize}
Intuitively, these actions can be explained as follows:
\begin{itemize}
\item
$f.m \at g$ is the action by which a process residing at locus $g$
requests a process residing at locus $f$ to carry out method $m$;
\item
$\passive g.m \at f$ is the action by which a process residing at locus
$f$ grants a request of a process residing at locus $g$ to carry out
method $m$;
\item
$f.m \uat g$ is the result of performing $f.m \at g$ and
$\passive g.m \at f$ at the same time.
\end{itemize}

In \ACPcc, the communication function
$\funct{\commm}{\Actd \x \Actd}{\Actd}$ is such that for all
$f,g \in \Loci$ and $m \in \Meth$:
\begin{itemize}
\item
$f.m \at g \commm \passive g.m \at f = f.m \uat g$;
\item
$f.m \at g \commm a = \dead$ for all
$a \in \Act \diff \set{\passive g.m \at f}$;
\item
$a \commm \passive g.m \at f = \dead$ for all
$a \in \Act \diff \set{f.m \at g}$;
\item
$f.m \uat g \commm a = \dead$ for all $a \in \Act$.
\end{itemize}

The receive actions and send actions commonly used for handshaking
communication of data, see e.g.~\cite{BW90}, can be viewed as requests
to carry out some communication method and grants of such requests,
respectively.
However, the current set-up requires that it is made explicit what are
the loci at which the sender and receiver involved reside.

\section{Integers}
\label{sect-INT}

In this section, we present an algebraic theory about integers which will
be used in later sections.
The presented theory is called \INT.

\INT\ has one sort: the sort $\Int$ of \emph{integers}.
To build terms of sort $\Int$, \INT\ has the following constants and
operators:
\begin{itemize}
\item
the constant $\const{0}{\Int}$;
\item
the constant $\const{1}{\Int}$;
\item
the binary \emph{addition} operator
$\funct{+}{\Int \x \Int}{\Int}$;
\item
the unary \emph{additive inverse} operator $\funct{-}{\Int}{\Int}$;
\item
the unary \emph{signum} operator $\funct{\sg}{\Int}{\Int}$.
\end{itemize}
Terms of sort $\Int$ are built as usual for a one-sorted signature.
Throughout the paper, we assume that there are infinitely many variables
of sort $\Int$, including $k$, $l$ and $n$.

As usual, we use infix notation for the binary operator ${} + {}$ and
prefix notation for the unary operator $-$.
The following additional precedence convention is used to reduce the
need for parentheses.
The operator ${} + {}$ binds weaker than the operator $-$.

The constants and operators of \INT\ are adopted from integer arithmetic
and need no further explanation.
The operator $\sg$ is useful where a distinction between positive
integers, negative integers and zero must be made.

The axioms of \INT\ are the axioms given in Table~\ref{axioms-INT}.%
\begin{table}[!t]
\caption{Axioms of \INT}
\label{axioms-INT}
\begin{eqntbl}
\begin{axcol}
0 + k = k                                           & \axiom{INT1} \\
-k + k =  0                                         & \axiom{INT2} \\
(k + l) + n = k + (l + n)                           & \axiom{INT3} \\
k + l = l + k                                       & \axiom{INT4}
\eqnsep
\sg(0) = 0                                          & \axiom{SG1} \\
\sg(1) = 1                                          & \axiom{SG2} \\
\sg(-1) = -1                                        & \axiom{SG3} \\
\sg(k + \sg(k)) = \sg(k)                            & \axiom{SG4}
\end{axcol}
\end{eqntbl}
\end{table}
Axioms INT1--INT4 are the axioms of a commutative group.
Axioms SG1--SG4 are the defining axioms of $\sg$.

The initial model of \INT\ is considered the standard model of \INT.

\section{Interface Group for Cooperating Components}
\label{sect-IFGcc}

In this section, we present an algebraic theory about interfaces.
The presented theory is called \IFGcc.
In Section~\ref{sect-ACC}, we will consider process components
which are taken as pairs of an interface and a process that is
made up of active actions, passive actions, and neutral actions.
Interfaces are built from two kinds of interface elements.

The set of \emph{interface elements} consists of:
\begin{itemize}
\item
for each $f,g \in \Loci$ and $m \in \Meth$,
the \emph{active interface element} $f.m \at g$;
\item
for each $f,g \in \Loci$ and $m \in \Meth$,
the \emph{passive interface element} $\passive f.m \at g$.
\end{itemize}
We write $\IFElems$ for the set of all interface elements.

Obviously, $\IFElems$ is a proper subset of $\Act$.
The interface elements are those actions that are allowed to occur in
interfaces.
The interface part of a process component consists of the active and
passive actions that the process part of that process component may be
capable of performing.
The interface elements $f.m \at g$ and $\passive g.m \at f$ are
considered each other inverses.
That is, if both occur in an interface, they cancel out.

Active interface elements are usually included in the interface of a
process component to couch that it expects from the environment in which
it is put the ability to grant certain requests.
Passive interface elements are usually included in the interface of a
process component to couch that it promises the environment in which it
is put the ability to make certain requests.
The environment in which the process component is put may be composed of
different process components.
To couch that it expects from a number of process components an ability
or it promises a number of process components an ability, the relevant
interface element is included the number of times concerned in the
interface of the process component.
An example of the need for multiple occurrences of interface elements in
interfaces of process components is found in
Section~\ref{sect-another-example}.

The distinction between active interface elements and passive interface
elements made here is a case of distinction between expectations and
promises because interface elements are actions that process components
may be capable of performing.
If the interface elements would be actions that process components must
be capable of performing, it would be a case of distinction between
requirements and provisions.

Interfaces can be considered multisets over the set of all active
interface elements in which multiplicities of elements may be negative
too, since occurrences of passive interface elements in an interface can
be counted as negative occurrences of their inverses.

\IFGcc\ has the sort $\Int$ from $\INT$ and in addition the sort $\IF$
of \emph{interfaces}.
To build terms of sort $\IF$, \IFGcc\ has the following constants and
operators:
\begin{itemize}
\item
the \emph{empty interface} constant $\const{0}{\IF}$;
\item
for each $e \in \IFElems$, the \emph{interface element} constant
$\const{e}{\IF}$;
\item
the binary \emph{interface combination} operator
$\funct{+}{\IF \x \IF}{\IF}$;
\item
the unary \emph{interface inversion} operator $\funct{-}{\IF}{\IF}$.
\end{itemize}
To build terms of sort $\Int$, \IFGcc\ has the constants and operators
of \INT\ and in addition the following operator:
\begin{itemize}
\item
for each $f,g \in \Loci$ and $m \in \Meth$, the unary
\emph{multiplicity} operator
$\funct{\mult{f.m \at g}}{\linebreak[2]\IF}{\Int}$.
\end{itemize}
Terms of the sorts $\IF$ and $\Int$ are built as usual for a many-sorted
signature (see e.g.~\cite{Wir90a,ST99a}).
Throughout the paper, we assume that there are infinitely many variables
of sort $\IF$, including $i$, $j$ and $h$.

We use infix notation for the binary operator $+$ and prefix notation
for the unary operator $-$.
The following precedence convention is used to reduce the need for
parentheses.
The operator ${} + {}$ binds weaker than the operator $-$.

Let $I$ and $J$ be closed terms of sort $\IF$, $f,g \in \Loci$, and
$m \in \Meth$.
Viewing interfaces as multisets with multiplicities from $\Int$, the
constants and operators of \IFGcc\ to build terms of sort $\IF$ can be
explained as follows:
\begin{itemize}
\item
$0$ is the interface in which the multiplicity of each active interface
element is $0$;
\item
$f.m \at g$ is the interface in which the multiplicity of $f.m \at g$ is
$1$ and the multiplicity of each other active interface element is $0$;
\item
$\passive f.m \at g$ is the interface in which the multiplicity of
$g.m \at f$ is $-1$ and the multiplicity of each other active interface
element is $0$;
\item
$I + J$ is the interface in which the multiplicity of each active
interface element is the addition of its multiplicities in $I$ and $J$;
\item
$-I$ is the interface in which the multiplicity of each active interface
element is the additive inverse of its multiplicity in $I$.
\end{itemize}
The operators $\mult{f.m \at g}$, one for each $f,g \in \Loci$ and
$m \in \Meth$, can simply be explained as follows:
\begin{itemize}
\item
$\mult{f.m \at g}(I)$ is the multiplicity of $f.m \at g$ in $I$.
\end{itemize}

We write $\Comb{i \in \cI} I_i$, where $\cI = \set{i_1,\ldots,i_n}$ and
$I_{i_1},\ldots,I_{i_n}$ are terms of sort $\IF$, for
$I_{i_1} + \ldots + I_{i_n}$.
The convention is that $\Comb{i \in \cI} I_i$ stands for $0$ if
$\cI = \emptyset$.

The axioms of \IFGcc\ are the axioms of \INT\ and the axioms given in
Table~\ref{axioms-IFG}.%
\begin{table}[!t]
\caption{Axioms of \IFGcc}
\label{axioms-IFG}
\begin{eqntbl}
\begin{axcol}
0 + i = i                                          & \axiom{IFG1} \\
-i + i = 0                                         & \axiom{IFG2} \\
(i + j) + h = i + (j + h)                          & \axiom{IFG3} \\
i + j = j + i                                      & \axiom{IFG4} \\
f.m \at g + \passive g.m \at f = 0                 & \axiom{IFG5}
\eqnsep
\mult{f.m \at g}(0) = 0                            & \axiom{M1} \\
\mult{f.m \at g}(f'.m' \at g') = 0
 \quad \mif f \neq f' \Or m \neq m' \Or g \neq g'  & \axiom{M2} \\
\mult{f.m \at g}(f.m \at g) = 1                    & \axiom{M3} \\
\mult{f.m \at g}(-i) = - \mult{f.m \at g}(i)       & \axiom{M4} \\
\mult{f.m \at g}(i + j) =
\mult{f.m \at g}(i) + \mult{f.m \at g}(j)          & \axiom{M5}
\end{axcol}
\end{eqntbl}
\end{table}
IFG5 and M1--M5 are actually axiom schemas in which $f$ and $g$ stand
for arbitrary members of $\Loci$ and $m$ stands for an arbitrary member
of $\Meth$.
Axioms IFG1--IFG4 are the axioms of a commutative group and axiom IFG5,
called the \emph{reflection law}, states that $\passive g.m \at f$ is
taken as the inverse of $f.m \at g$.
Axioms M1--M5 are the defining axioms of $\mult{f.m \at g}$.

The initial model of \IFGcc\ is considered the standard model of \IFGcc.

Other interface groups for cooperating components are conceivable.
For example, adding $i + i = 0$, or equivalently $i = - i$, to the
axioms of \IFGcc\ yields an interface group with torsion.
This addition means that no distinction is made between an active
interface element and the passive interface element that is its inverse.
This is not unfamiliar.
\IFGcc\ without torsion goes with the observable actions of
CCS~\cite{Mil89}, whereas \IFGcc\ with torsion goes with the events of
CSP~\cite{Hoa85}.

\section{Algebra of Cooperating Components}
\label{sect-ACC}

In this section, we take up the extension of \ACPcc\ to a theory about
process components.
The result is called \ACC\ (Algebra of Cooperating Components).

In the preceding sections, we have already been gone into some of the
general ideas that underlie the design of this extension.
Those ideas, which concern the interfaces and behaviours of process
components, can be summarized as follows:
\begin{itemize}
\item
behaviours of process components are processes made up of three kinds of
actions: active actions, passive actions and neutral actions;
\item
for each active action, there is a unique passive action with which it
can be performed synchronously, and vice versa;
\item
interfaces of process components consist of active and passive actions
that the process components may be capable of performing;
\item
looked upon as an interface element, each active action has the passive
action with which it can be performed synchronously as its inverse, and
vice versa;
\item
in interfaces of process components, there may be elements with multiple
occurrences.
\end{itemize}
The remaining general ideas concern the process components by
themselves:
\begin{itemize}
\item
if a process is turned into a process component by adding an interface
to it, the process is restricted by the interface with respect to the
active and passive actions that it can perform to force that the
behaviour of the process component complies with its interface;
\item
if two process components are composed, the interface of the composed
process component is the combination of the interfaces of the two
process components and the behaviour of the composed process component
is the parallel composition of the behaviours of the two process
components restricted by the combination of the interfaces of the two
process components.
\end{itemize}

The point of view on the composition of process components implies that,
if all occurrences of an (active or passive) action in the interface of
a process component are cancelled out by composition with another
process component, this action is blocked in the behaviour of the
composition of these process components.
The blocking of the action takes place even if its inverse is not
included in the actions that make up the behaviour of the other process
component.
It is possible that the inverse is not included because the interfaces
concern expectations and promises instead of requirements and
provisions (see also Section~\ref{sect-IFGcc}).
The way in which is dealt with this possibility can be explained as
follows:
(i)~if a promised ability to make a request is not provided, making the
request is blocked and
(ii)~if an expected ability to grant a request is not required, granting
the request is blocked.

\ACC\ has the sort $\Proc$ from \ACPcc, the sorts $\IF$ and $\Int$ from
\IFGcc, and in addition the sort $\Comp$ of \emph{components}.
To build terms of sort $\Comp$, \ACC\ has the following operators:
\begin{itemize}
\item
the binary \emph{basic component} operator
$\funct{\comp}{\IF \x \Proc}{\Comp}$;
\item
the binary \emph{component composition} operator
$\funct{\parc}{\Comp \x \Comp}{\Comp}$.
\end{itemize}
To build terms of sort $\Proc$, \ACC\ has the constants and operators of
\ACPcc\ and in addition the following operator:
\begin{itemize}
\item
the binary \emph{interface compliant encapsulation} operator
$\funct{\iencap{}}{\IF \x \Proc}{\Proc}$.
\end{itemize}
To build terms of sort $\IF$, \ACC\ has the constants and operators of
\IFGcc\ to build terms of sort $\IF$.
To build terms of sort $\Int$, \ACC\ has the constants and operators of
\IFGcc\ to build terms of sort $\Int$.

Terms of the different sorts are built as usual for a many-sorted
signature.
Throughout the paper, we assume that there are infinitely many variables
of sort $\Comp$, including $u$, $v$, $u'$ and $v'$.

We use infix notation for the binary operator $\parc$.
We write $\iencap{I}(P)$, where $I$ is a term of sort $\IF$ and
$P$ is a term of sort $\Proc$, for $\iencap{{}}(I,P)$.

Let $C$ and $D$ be closed terms of sort $\Comp$, $P$ be a closed term of
sort $\Proc$, and $I$ be a closed term of sort $\IF$.
Viewing interfaces as multisets with multiplicities from $\Int$, the
operators of \ACC\ to build terms of sort $\Comp$ can be explained as
follows:
\begin{itemize}
\item
$\comp(I,P)$ is the process component of which the interface is $I$ and
the behaviour is $P$, except that active actions of which the
multiplicity in $I$ is not positive and passive actions with an inverse
of which the multiplicity in $I$ is not negative are blocked;
\item
$C \parc D$, is the process component of which the interface is the
combination of the interfaces of $C$ and $D$ and the behaviour is the
parallel composition of the behaviours of $C$ and $D$, except that
active actions of which the multiplicity in the combination of the
interfaces of $C$ and $D$ is not positive and passive actions with an
inverse of which the multiplicity in the combination of the interfaces
of $C$ and $D$ is not negative are blocked.
\end{itemize}
The operator $\iencap{{}}$ can be explained as follows:
\begin{itemize}
\item
$\iencap{I}(P)$ behaves the same as $P$, except that active actions of
which the multiplicity in $I$ is not positive and passive actions with
an inverse of which the multiplicity in $I$ is not negative are blocked.
\end{itemize}
The operator $\iencap{{}}$ is an auxiliary operator used in the axioms
concerning process components.

The axioms of \ACC\ are the axioms of \ACP, the axioms of \IFGcc, and
the axioms given in Table~\ref{axioms-ACC}.%
\begin{table}[!t]
\caption{Axioms of \ACC}
\label{axioms-ACC}
\begin{eqntbl}
\begin{axcol}
\comp(i,x) = \comp(i,\iencap{i}(x))                     & \axiom{CC1} \\
\comp(i,x) \parc \comp(j,y) =
\comp(i + j,\iencap{i}(x) \parc \iencap{j}(y))          & \axiom{CC2}
\eqnsep
\sg(\mult{f.m \at g}(i)) =  1 \Implies
                      \iencap{i}(f.m \at g) = f.m \at g & \axiom{E1} \\
\sg(\mult{f.m \at g}(i)) =  0 \Implies
                          \iencap{i}(f.m \at g) = \dead & \axiom{E2} \\
\sg(\mult{f.m \at g}(i)) = -1 \Implies
                          \iencap{i}(f.m \at g) = \dead & \axiom{E3} \\
\sg(\mult{g.m \at f}(i)) = -1 \Implies
    \iencap{i}(\passive f.m \at g) = \passive f.m \at g & \axiom{E4} \\
\sg(\mult{g.m \at f}(i)) =  0 \Implies
                 \iencap{i}(\passive f.m \at g) = \dead & \axiom{E5} \\
\sg(\mult{g.m \at f}(i)) =  1 \Implies
                 \iencap{i}(\passive f.m \at g) = \dead & \axiom{E6} \\
\iencap{i}(f.m \uat g) = f.m \uat g                     & \axiom{E7} \\
\iencap{i}(\dead) = \dead                               & \axiom{E8} \\
\iencap{i}(x \altc y) = \iencap{i}(x) \altc \iencap{i}(y)
                                                        & \axiom{E9} \\
\iencap{i}(x \seqc y) = \iencap{i}(x) \seqc \iencap{i}(y)
                                                        & \axiom{E10}
\end{axcol}
\end{eqntbl}
\end{table}
E1--E7 are actually axiom schemas in which $f$ and $g$ stand for
arbitrary members of $\Loci$ and $m$ stands for an arbitrary member of
$\Meth$.
Axioms CC1 and CC2 are axioms concerning process components and axioms
E1--E10 are the defining axioms of the auxiliary operator $\iencap{{}}$.
Together they formalize the intuition about process components given
above in a direct way.
It is only because they are used in axioms E1--E6 that the multiplicity
operators $\mult{f.m \at g}$ are included in the theory \IFGcc\ and the
signum operator $\sg$ is included in the theory \INT.

Guarded recursion can be added to \ACC\ as it is added to \ACP\ in
Section~\ref{sect-REC}.
We write \ACC+\REC\ for \ACC\ extended with the constants standing for
the unique solutions of guarded recursive specifications and the axioms
RDP and RSP.

In Section~\ref{sect-model}, we will construct a model of \ACC+\REC\
using a notion of bisimilarity for process components.

\section{On the Associativity of Component Composition}
\label{sect-non-assoc}

In this section, we show that component composition is in general not
associative and couch in a special axiom that component composition is
associative when a certain condition on its operands is fulfilled.

Let $f,g \in \Loci$, and let $m,m',m'' \in \Meth$ be such that
$m' \neq m''$, and take
\begin{ldispl}
C_1 = \comp(\passive g.m \at f + g.m' \at f,
            \passive g.m \at f \seqc g.m' \at f)\;, \\
C_2 = \comp(f.m \at g,f.m \at g)\;, \\
C_3 = \comp(\passive g.m \at f + g.m'' \at f,
            \passive g.m \at f \seqc g.m'' \at f)\;.
\end{ldispl}
We easily derive from the axioms of \ACC\ that
\begin{ldispl}
(C_1 \parc C_2) \parc C_3 = {} \\
\comp(g.m' \at f,f.m \uat g \seqc g.m' \at f) \parc C_3 = {} \\
\comp(\passive g.m \at f + g.m' \at f + g.m'' \at f,
       f.m \uat g \seqc g.m' \at f \seqc \dead)
\end{ldispl}
and
\begin{ldispl}
C_1 \parc (C_2 \parc C_3) = {} \\
C_1 \parc \comp(g.m'' \at f,f.m \uat g \seqc g.m'' \at f) = {} \\
\comp(\passive g.m \at f + g.m' \at f + g.m'' \at f,
       f.m \uat g \seqc g.m'' \at f \seqc \dead)\;.
\end{ldispl}
Hence, we have that
$(C_1 \parc C_2) \parc C_3 \neq C_1 \parc (C_2 \parc C_3)$.

The associativity axiom for component composition is given
in Table~\ref{axioms-assoc-comp}.%
\begin{table}[!t]
\caption{Associativity axiom for component composition}
\label{axioms-assoc-comp}
\begin{eqntbl}
\begin{eqncol}
\AND{f,g \in \Loci,m \in \Meth}
(\mult{f.m \at g}(i + j + h) = 0 \Or {}
\\ \phantom{\AND{f,g \in \Loci,m \in \Meth} (}\!
 \mult{f.m \at g}(i) = 0 \Or \mult{f.m \at g}(j) = 0 \Or
 \mult{f.m \at g}(h) = 0 \Or {}
\\ \phantom{\AND{f,g \in \Loci,m \in \Meth} (}\!
 \sg(\mult{f.m \at g}(i)) = \sg(\mult{f.m \at g}(j)) \And
 \sg(\mult{f.m \at g}(j)) = \sg(\mult{f.m \at g}(h)))
 \Implies \\
(\comp(i,x) \parc \comp(j,y)) \parc \comp(h,z) =
 \comp(i,x) \parc (\comp(j,y) \parc \comp(h,z))
\end{eqncol}
\end{eqntbl}
\end{table}
It is not known to us whether the condition in this axiom is a necessary
condition for associativity of component composition.

Below, we will sketch the justification of the associativity axiom.
For that purpose, we first shortly introduce the approximation induction
principle, which has been introduced before in the setting of \ACP.

Guarded recursion gives rise to infinite processes.
In \ACC+\REC, closed terms of sort $\Proc$ that denote the same infinite
process cannot always be proved equal by means of the axioms of
\ACC+\REC.
To remedy this, we can add the approximation induction principle to
\ACC+\REC.
The approximation induction principle, \AIP\ in short, was first
formulated in the setting of \ACP\ in~\cite{BK86c}.
It formalized the idea that two processes are identical if their
approximations up to any finite depth are identical.
The approximation up to depth $n$ of a process behaves the same as that
process, except that it cannot perform any further action after $n$
actions have been performed.
Approximation up to depth $n$ is phrased in terms of the unary
\emph{projection} operator $\proj{n}$.
For a comprehensive treatment of projections and AIP, the reader is
referred to~\cite{BW90}.

We proceed with the justification of the associativity axiom given in
Table~\ref{axioms-assoc-comp}.
It can be proved that all closed substitution instances of this axiom
are derivable from the axioms of \ACC+\REC, the axioms for the
projection operators and \AIP.
Moreover, the model $\gB_{\ACC+\REC}$ of \ACC+\REC\ that will be
constructed in Section~\ref{sect-model} can be expanded with operations
for the projection operators such that the axioms for the projection
operators and \AIP\ hold in the expansion.
Because all elements of the sets associated with the sorts $\Proc$,
$\IF$ and $\Comp$ in $\gB_{\ACC+\REC}$ are interpretations of closed
terms, it follows that the associativity axiom holds in
$\gB_{\ACC+\REC}$.

\section{Closed Systems and Interfaces of Process Components}
\label{sect-closed-systems}

In this short section, we discuss the connection between closed systems
and empty interfaces.
The intuition is that a system is a closed system if the actions that
make up its behaviour include neither active actions nor passive
actions.

We first shortly introduce the alphabet operator, which has been
introduced before in the setting of \ACP.

The set of actions that can be performed by a process is called the
alphabet of the process.
We can add the unary \emph{alphabet} operator $\alpha$ to \ACC+\REC\ to
extract the alphabet from a process.
The alphabet operator was first added to \ACP+\REC\ in~\cite{BBK87a}.
To deal with infinite processes, the projection operators occur in the
axioms for this operator.
For a comprehensive treatment of alphabets, the reader is referred
to~\cite{BW90}.

Let $I$ be a closed term of sort $\IF$ and $P$ be a closed term of sort
$\Proc$.
Then $\comp(I,P)$ is a \emph{closed system} if
$\alpha(\iencap{I}(P)) \subseteq
 \set{f.m \uat g \where f,g \in \Loci, m \in \Meth}$.

It can be proved that, for each closed term $I$ of sort $\IF$ and closed
term  $P$ of sort $\Proc$, the following is derivable from the axioms of
\ACC+\REC, the axioms for the alphabet operator, the axioms for the
projection operators and \AIP:
\begin{ldispl}
I = 0 \Implies \comp(I,P) \mathrm{\;is\; a\; closed\; system}\;.
\end{ldispl}
It is generally undecidable whether $\comp(I,P)$ is a closed system.
However, it is decidable whether $I = 0$.
This illustrates the usefulness combining a process with an interface in
the way presented in this paper.

\section{An Example}
\label{sect-example}

In this section, we illustrate the use of \ACC\ by means of an example
concerning buffers with capacity one.
We assume a finite set $\cD$ of data with $\err \in \cD$ and, for each
$d \in \cD$, a method $c_d$ for communicating datum $d$.
We take the element $\err \in \cD$ for an improper datum.

We consider the three buffer processes $B_f$, $B_g$, and $B_h$ that are
defined by the guarded recursion equations
\begin{ldispl}
\begin{aeqns}
B_f & = &
\Altcv{d \in \cD \diff \set{\err}}
 \passive s.c_d \at f \seqc
 (g.c_d \at f \altc g.c_\err \at f) \seqc B_f\;,
\\
B_g & = &
\Altcv{d \in \cD \diff \set{\err}}
 \passive f.c_d \at g \seqc
 (h.c_d \at g \altc h.c_\err \at g) \seqc B_g\;,
\\
B_h & = &
\Altcv{d \in \cD \diff \set{\err}}
 \passive g.c_d \at h \seqc
 (r.c_d \at h \altc r.c_\err \at h) \seqc B_h\;,
\end{aeqns}
\end{ldispl}
respectively.
The processes $B_f$, $B_g$ and $B_h$ always reside at the loci $f$, $g$
and $h$, respectively.
$B_f$ is able to pass data from a process residing at locus $s$ to a
process residing at locus $g$, $B_g$ is able to pass data from a process
residing at locus $f$ to a process residing at locus $h$, and $B_h$ is
able to pass data from a process residing at locus $g$ to a process
residing at locus $r$.
$B_f$, $B_g$ and $B_h$ are faulty in the sense that they may deliver an
improper datum instead of the datum to be delivered.

We turn these three buffer processes into process components by adding
interfaces to them.
To be exact, we turn the processes $B_f$, $B_g$, and $B_h$ into the
process components $\comp(I_f,B_f)$, $\comp(I_g,B_g)$, and
$\comp(I_h,B_h)$, where
\begin{ldispl}
\begin{aeqns}
I_f & = &
\Combv{d \in \cD \diff \set{\err}} \passive s.c_d \at f +
\Combv{d \in \cD} g.c_d \at f\;,
\\
I_g & = &
\Combv{d \in \cD \diff \set{\err}} \passive f.c_d \at g +
\Combv{d \in \cD} h.c_d \at g\;,
\\
I_h & = &
\Combv{d \in \cD \diff \set{\err}} \passive g.c_d \at h +
\Combv{d \in \cD} r.c_d \at h\;.
\end{aeqns}
\end{ldispl}

We have a look at the component composition
$\comp(I_f,B_f) \parc (\comp(I_g,B_g) \parc \comp(I_h,B_h))$
--~which equals
$(\comp(I_f,B_f) \parc \comp(I_g,B_g)) \parc \comp(I_h,B_h)$
by the associativity axiom for component composition.
It follows from axioms CC1 and CC2 that
\begin{ldispl}
\comp(I_f,B_f) \parc (\comp(I_g,B_g) \parc \comp(I_h,B_h))
\\ \quad {} =
\comp\bigl(I_f + I_g + I_h,
           \iencap{I_f + I_g + I_h}
            (\iencap{I_f}(B_f) \parc
             \iencap{I_g + I_h}
              (\iencap{I_g}(B_g) \parc \iencap{I_h}(B_h)))\bigr)\;.
\end{ldispl}
Moreover, it follows from axioms IFG1--IFG5 that
\begin{ldispl}
I_f + I_g + I_h =
\Combv{d \in \cD \diff \set{\err}} \passive s.c_d \at f +
g.c_\err \at f + h.c_\err \at g +
\Combv{d \in \cD} r.c_d \at h
\end{ldispl}
and from axioms INT1--INT4, SG1--SG4, IFG5, M1--M5, E1--E10, and RSP
that
\begin{ldispl}
\iencap{I_f + I_g + I_h}
 (\iencap{I_f}(B_f) \parc
  \iencap{I_g + I_h}(\iencap{I_g}(B_g) \parc \iencap{I_h}(B_h)))
\\ \quad {} =
\iencap{I_f + I_g + I_h}(B_f \parc B_g \parc B_h)\;.
\end{ldispl}
Hence, we have by axiom CC1 that
\begin{ldispl}
\comp(I_f,B_f) \parc (\comp(I_g,B_g) \parc \comp(I_h,B_h))
\\ \quad {} =
\comp\Biggl(\Combv{d \in \cD \diff \set{\err}} \passive s.c_d \at f +
            g.c_\err \at f + h.c_\err \at g +
            \Combv{d \in \cD} r.c_d \at h,
            B_f \parc B_g \parc B_h\Biggr)\;.
\end{ldispl}
It can further be shown by means of the axioms of \ACP+\REC\ that the
behaviour of
$\comp(I_f,B_f) \parc (\comp(I_g,B_g) \parc \comp(I_h,B_h))$
is essentially a buffer with capacity three.
This buffer process, which resides alternately at the loci $f$, $g$ and
$h$, is able to pass data from a process residing at locus $s$ to a
process residing at locus $r$.
It is faulty in the sense that it may deliver an improper datum instead
of the datum to be delivered.
Moreover, the improper datum may be delivered at the locus $g$ or the
locus $h$ instead of the locus $r$.

The process component
$\comp(I_f,B_f) \parc (\comp(I_g,B_g) \parc \comp(I_h,B_h))$
does not have an empty interface.
It follows from axioms IFG1--IFG5 that composing it with a process
component whose interface is
\begin{ldispl}
\Combv{d \in \cD \diff \set{\err}} f.c_d \at s +
\passive f.c_\err \at g + \passive g.c_\err \at h +
\Combv{d \in \cD} \passive h.c_d \at r
\end{ldispl}
would result in an empty interface.
This shows that an empty interface requires composition with a process
component that promises to handle the delivery of an improper datum at
the loci $g$, $h$ and $r$.

\section{Another Example}
\label{sect-another-example}

In this section, we illustrate the use of \ACC\ by means of an example
in which a single buffer with capacity one is used to pass data between
three components.
We assume a finite set $\cD$ of data, a function $\funct{F}{\cD}{\cD}$
and, for each $d \in \cD$, a method $c_d$ for communicating datum $d$.
We also assume methods $\nm{wa}_1$, $\nm{wa}_2$, $\nm{wa}_3$,
$\nm{sl}_1$, $\nm{sl}_2$ and $\nm{sl}_3$ for controlling the cooperation
of the three components that share the buffer.

We consider the processes $P_1$, $P_2$ and $P_3$ that are defined by the
guarded recursion equations
\begin{ldispl}
\begin{aeqns}
P_1 & = &
\passive h.\nm{wa}_1 \at f \seqc
\Altcv{d \in \cD}
 \passive s.c_d \at f \seqc g.c_d \at f \seqc
 \passive h.\nm{sl}_1 \at f \seqc P_1\;,
\\
P_2 & = &
\passive h.\nm{wa}_2 \at f \seqc
\Altcv{d \in \cD}
 \passive g.c_d \at f \seqc g.c_{F(d)} \at f \seqc
 \passive h.\nm{sl}_2 \at f \seqc P_2\;,
\\
P_3 & = &
\passive h.\nm{wa}_3 \at f \seqc
\Altcv{d \in \cD}
 \passive g.c_d \at f \seqc r.c_d \at f \seqc
 \passive h.\nm{sl}_3 \at f \seqc P_3\;,
\end{aeqns}
\end{ldispl}
respectively.
All three processes always reside at locus $f$.
$P_1$ is able to pass data from a process residing at locus $s$ to a
process residing at locus $g$, $P_2$ is able to apply an operation to
data hold by a process residing at locus $g$, and $P_3$ is able to pass
data from a process residing at locus $g$ to a process residing at locus
$r$.
The processes $P_1$, $P_2$ and $P_3$ are called the entry process, the
main process and the exit process, respectively.
We also consider the buffer process $B$ and the control process $C$
defined by the guarded recursion equations
\begin{ldispl}
\begin{aeqns}
B & = &
\Altcv{d \in \cD}
 \passive f.c_d \at g \seqc f.c_d \at g \seqc B\;,
\\
C & = &
\Altcv{d \in \cD}
 f.\nm{wa}_1 \at h \seqc f.\nm{sl}_1 \at h \seqc
 f.\nm{wa}_2 \at h \seqc f.\nm{sl}_2 \at h \seqc
 f.\nm{wa}_3 \at h \seqc f.\nm{sl}_3 \at h \seqc C\;,
\end{aeqns}
\end{ldispl}
respectively.
The processes $B$ and $C$ always reside at the loci $g$ and $h$,
respectively.
$B$ is able to pass data from a process residing at locus $f$ to
a process residing at locus $f$ and
$C$ is able to control the cooperation of three processes residing at
locus $f$ such that they take turns in doing a number of steps.

We turn all these processes into process components by adding interfaces
to them.
To be exact, we turn $P_1$, $P_2$, $P_3$, $B$ and $C$ into the process
components $\comp(I_1,P_1)$, $\comp(I_2,P_2)$, $\comp(I_3,P_3)$,
$\comp(J,B)$ and $\comp(H,C)$, where
\begin{ldispl}
\begin{aeqns}
I_1 & = &
\Combv{d \in \cD}
 (\passive s.c_d \at f + g.c_d \at f) +
\passive h.\nm{wa}_1 \at f + \passive h.\nm{sl}_1 \at f\;,
\\
I_2 & = &
\Combv{d \in \cD}
 (\passive g.c_d \at f + g.c_d \at f) +
\passive h.\nm{wa}_2 \at f + \passive h.\nm{sl}_2 \at f\;,
\\
I_3 & = &
\Combv{d \in \cD}
 (\passive g.c_d \at f + r.c_d \at f) +
\passive h.\nm{wa}_3 \at f + \passive h.\nm{sl}_3 \at f\;,
\\
J & = &
\Combv{d \in \cD}
 (\passive f.c_d \at g + \passive f.c_d \at g +
  f.c_d \at g + f.c \at g)\;,
\\
H & = &
 f.\nm{wa}_1 \at h + f.\nm{wa}_2 \at h + f.\nm{wa}_3 \at h +
 f.\nm{sl}_1 \at h + f.\nm{sl}_2 \at h + f.\nm{sl}_3 \at h\;.
\end{aeqns}
\end{ldispl}
Notice that $g.c_d \at f$ occurs once in both $I_1$ and $I_2$ and
$\passive g.c_d \at f$ occurs once in both $I_2$ and $I_3$, whereas
their inverses occur twice in $J$.

We have a look at
$\comp(I_1,P_1) \parc (\comp(I_2,P_2) \parc (\comp(I_3,P_3) \parc
 (\comp(J,B) \parc \comp(H,C))))$.\linebreak[2]
It follows from the axioms of \ACC+\REC\ that
\begin{ldispl}
\comp(I_1,P_1) \parc (\comp(I_2,P_2) \parc (\comp(I_3,P_3) \parc
 (\comp(J,B) \parc \comp(H,C))))
\\ \quad {} =
\comp\Biggl(\Combv{d \in \cD} (\passive s.c_d \at f + r.c_d \at f),
            P_1 \parc P_2 \parc P_3 \parc B \parc C\Biggr)\;.
\end{ldispl}
This would not be case if $\passive f.c_d \at g$ and $f.c_d \at g$ would
occur only once in $J$.
The behaviour of
$\comp(I_1,P_1) \parc (\comp(I_2,P_2) \parc (\comp(I_3,P_3) \parc
 (\comp(J,B) \parc \comp(H,C))))$
is essentially a process that is able to receive data from a process
residing at locus $s$, apply $F$ to the received data, and send the
results to a process residing at locus $r$.
Each cycle of the process is accomplished as follows: first $P_1$
receives a datum and puts it in buffer $B$, then $P_2$ gets the datum
from buffer $B$, applies $F$ to it and put the result back in buffer
$B$, and finally $P_3$ gets the result from buffer $B$ and sends the
result.
$C$ controls that $P_1$, $P_2$ and $P_3$ do not start their part of the
cycle prematurely.

\section{Bisimilarity of Process Components}
\label{sect-bisim}

In this section, we give a structural operational semantics for
\ACC+\REC\ and define a notion of bisimilarity based on it.
This notion of bisimilarity will be used in Section~\ref{sect-model} to
construct a model of \ACC+\REC.

Henceforth, we will write $\cT_S$, where
$S \in \set{\Proc,\IF,\Comp}$, for the set of all closed terms of
sort $S$ from the language of \ACC+\REC.
Moreover, we will write $\cT^\sINT_\Int$ for the set of all closed terms
of sort $\Int$ from the language of \INT.

The following relations are the primary relations used in the structural
operational semantics of \ACC+\REC:
\begin{itemize}
\item
a unary relation ${\termp{a}} \subseteq \cT_\Proc$,
for each $a \in \Act$;
\item
a binary relation
${\stepp{a}} \subseteq \cT_\Proc \x \cT_\Proc$,
for each $a \in \Act$;
\item
a unary relation ${\inm{f.m \at g\!}{N}{}} \subseteq \cT_\IF$,
for each $f,g \in \Loci$, $m \in \Meth$ and $N \in \cT^\sINT_\Int$;
\item
a binary relation ${\hasif{}{}} \subseteq \cT_\Comp \x \cT_\IF$;
\item
a unary relation ${\termc{a}} \subseteq \cT_\Comp$,
for each $a \in \Act$;
\item
a binary relation
${\stepc{a}} \subseteq \cT_\Comp \x \cT_\Comp$,
for each $a \in \Act$.
\end{itemize}
We write
$\atermp{P}{a}$ instead of $P \in {\termp{a}}$,
$\astepp{P}{a}{P'}$ instead of $\tup{P,P'} \in {\stepp{a}}$,
$\inm{f.m \at g}{N}{I}$ instead of
$I \in {\inm{f.m \at g\!}{N}{}}$,
$\hasif{C}{I}$ instead of $\tup{C,I} \in {\hasif{}{}}$,
$\atermc{C}{a}$ instead of $C \in {\termc{a}}$, and
$\astepc{C}{a}{C'}$ instead of $\tup{C,C'} \in {\stepc{a}}$.
The relations can be explained as follows:
\begin{itemize}
\item
$\atermp{P}{a}$:
process $P$ is capable of first performing $a$ and then terminating
successfully;
\item
$\astepp{P}{a}{P'}$:
process $P$ is capable of first performing $a$ and then proceeding as
process $P'$;
\item
$\inm{f.m \at g}{N}{I}$:
$f.m \at g$ occurs $N$ times in interface $I$;
\item
$\hasif{C}{I}$:
the interface of component $C$ is $I$;
\item
$\atermc{C}{a}$:
component $C$ is capable of first performing $a$ and then terminating
successfully;
\item
$\astepc{C}{a}{C'}$:
component $C$ is capable of first performing $a$ and then proceeding as
component $C'$.
\end{itemize}

The following relations are auxiliary relations used in the structural
operational semantics of \ACC+\REC:
\begin{itemize}
\item
a unary relation ${\inm{f.m \at g\!}{+}{}} \subseteq \cT_\IF$,
for each $f,g \in \Loci$ and $m \in \Meth$;
\item
a unary relation ${\inm{f.m \at g\!}{-}{}} \subseteq \cT_\IF$,
for each $f,g \in \Loci$ and $m \in \Meth$;
\item
a unary relation $\inifprel{f.m \at g\!} \subseteq \cT_\Comp$,
for each $f,g \in \Loci$ and $m \in \Meth$;
\item
a unary relation $\inifnrel{f.m \at g\!} \subseteq \cT_\Comp$,
for each $f,g \in \Loci$ and $m \in \Meth$.
\end{itemize}
We write $\inm{f.m \at g}{+}{I}$ and $\inm{f.m \at g}{-}{I}$ instead of
$I \in {\inm{f.m \at g\!}{+}{}}$ and $I \in {\inm{f.m \at g\!}{-}{}}$,
respectively.
We write $\inifp{f.m \at g}{C}$ and $\inifn{f.m \at g}{C}$ instead of
$C \in \inifprel{f.m \at g\!}$ and $C \in \inifnrel{f.m \at g\!}$,
respectively.
The relations can be explained as follows:
\begin{itemize}
\item
$\inm{f.m \at g}{+}{I}$:
$f.m \at g$ occurs a positive number of times in interface $I$;
\item
$\inm{f.m \at g}{-}{I}$:
$f.m \at g$ occurs a negative number of times in interface $I$;
\item
$\inifp{f.m \at g}{C}$: $f.m \at g$ occurs a positive number of times in
the interface of component $C$;
\item
$\inifn{f.m \at g}{C}$: $f.m \at g$ occurs a negative number of times in
the interface of component $C$.
\end{itemize}
The auxiliary relations are for convenience only.

The structural operational semantics of \ACC+\REC\ is described by the
rules given in Tables~\ref{rules-ACP+REC} and~\ref{rules-ACC}.
\begin{table}[!t]
\caption{Rules for operational semantics of \ACP+\REC}
\label{rules-ACP+REC}
\begin{ruletbl}
\Rule
{}
{\atermp{a}{a}}
\quad
\\
\Rule
{\atermp{x}{a}}
{\atermp{x \altc y}{a}}
\quad
\Rule
{\atermp{y}{a}}
{\atermp{x \altc y}{a}}
\quad
\Rule
{\astepp{x}{a}{x'}}
{\astepp{x \altc y}{a}{x'}}
\quad
\Rule
{\astepp{y}{a}{y'}}
{\astepp{x \altc y}{a}{y'}}
\\
\Rule
{\atermp{x}{a}}
{\astepp{x \seqc y}{a}{y}}
\quad
\Rule
{\astepp{x}{a}{x'}}
{\astepp{x \seqc y}{a}{x' \seqc y}}
\\
\Rule
{\atermp{x}{a}}
{\astepp{x \parc y}{a}{y}}
\quad
\Rule
{\atermp{y}{a}}
{\astepp{x \parc y}{a}{x}}
\quad
\Rule
{\astepp{x}{a}{x'}}
{\astepp{x \parc y}{a}{x' \parc y}}
\quad
\Rule
{\astepp{y}{a}{y'}}
{\astepp{x \parc y}{a}{x \parc y'}}
\\
\RuleC
{\atermp{x}{a},\; \atermp{y}{b}}
{\atermp{x \parc y}{c}}
{a \commm b = c}
\quad
\RuleC
{\atermp{x}{a},\; \astepp{y}{b}{y'}}
{\astepp{x \parc y}{c}{y'}}
{a \commm b = c}
\\
\RuleC
{\astepp{x}{a}{x'},\; \atermp{y}{b}}
{\astepp{x \parc y}{c}{x'}}
{a \commm b = c}
\quad
\RuleC
{\astepp{x}{a}{x'},\; \astepp{y}{b}{y'}}
{\astepp{x \parc y}{c}{x' \parc y'}}
{a \commm b = c}
\\
\Rule
{\atermp{x}{a}}
{\astepp{x \leftm y}{a}{y}}
\quad
\Rule
{\astepp{x}{a}{x'}}
{\astepp{x \leftm y}{a}{x' \parc y}}
\\
\RuleC
{\atermp{x}{a},\; \atermp{y}{b}}
{\atermp{x \commm y}{c}}
{a \commm b = c}
\quad
\RuleC
{\atermp{x}{a},\; \astepp{y}{b}{y'}}
{\astepp{x \commm y}{c}{y'}}
{a \commm b = c}
\\
\RuleC
{\astepp{x}{a}{x'},\; \atermp{y}{b}}
{\astepp{x \commm y}{c}{x'}}
{a \commm b = c}
\quad
\RuleC
{\astepp{x}{a}{x'},\; \astepp{y}{b}{y'}}
{\astepp{x \commm y}{c}{x' \parc y'}}
{a \commm b = c}
\\
\RuleC
{\atermp{x}{a}}
{\atermp{\encap{H}(x)}{a}}
{a \not\in H}
\quad
\RuleC
{\astepp{x}{a}{x'}}
{\astepp{\encap{H}(x)}{a}{\encap{H}(x')}}
{a \not\in H}
\\
\RuleC
{\atermp{\rec{t_X}{E}}{a}}
{\atermp{\rec{X}{E}}{a}}
{X = t_X \in E}
\quad
\RuleC
{\astepp{\rec{t_X}{E}}{a}{x'}}
{\astepp{\rec{X}{E}}{a}{x'}}
{X = t_X \in E}
\end{ruletbl}
\end{table}
%
% !!
\begin{table}[!p]
\caption{Additional rules for operational semantics of \ACC+\REC}
\label{rules-ACC}
\begin{ruletbl}
\Rule
{}
{\inm{f.m \at g}{1}{f.m \at g}}
\quad
\RuleC
{}
{\inm{f.m \at g}{0}{f'.m' \at g'}}
{f \neq f' \Or m \neq m' \Or g \neq g'}
\\
\Rule
{}
{\inm{f.m \at g}{-1}{\passive g.m \at f}}
\quad
\RuleC
{}
{\inm{f.m \at g}{0}{\passive g'.m' \at f'}}
{f \neq f' \Or m \neq m' \Or g \neq g'}
\\
\Rule
{}
{\inm{f.m \at g}{0}{0}}
\quad
\Rule
{\inm{f.m \at g}{k}{i}}
{\inm{f.m \at g}{-k}{-i}}
\quad
\Rule
{\inm{f.m \at g}{k}{i},\, \inm{f.m \at g}{l}{j}}
{\inm{f.m \at g}{k+l}{i + j}}
\\
\Rule
{\phantom{\hasif{\comp(i,x)}{i}}}
{\hasif{\comp(i,x)}{i}}
\quad
\Rule
{\hasif{u}{i},\, \hasif{v}{j}}
{\hasif{u \parc v}{i + j}}
\\
\Rule
{\inm{f.m \at g}{k}{i},\, \sg(k) = 1}
{\inm{f.m \at g}{+}{i}}
\quad
\Rule
{\inm{f.m \at g}{k}{i},\, \sg(k) = -1}
{\inm{f.m \at g}{-}{i}}
\\
\Rule
{\hasif{u}{i},\, \inm{f.m \at g}{+}{i}}
{\inifp{f.m \at g}{u}}
\quad
\Rule
{\hasif{u}{i},\, \inm{f.m \at g}{-}{i}}
{\inifn{f.m \at g}{u}}
\\
\Rule
{\atermp{x}{f.m \at g},\, \inm{f.m \at g}{+}{i}}
{\atermc{\comp(i,x)}{f.m \at g}}
\quad
\Rule
{\atermp{x}{\passive f.m \at g},\, \inm{g.m \at f}{-}{i}}
{\atermc{\comp(i,x)}{\passive f.m \at g}}
\quad
\Rule
{\atermp{x}{f.m \uat g}}
{\atermc{\comp(i,x)}{f.m \uat g}}
\\
\Rule
{\astepp{x}{f.m \at g}{x'},\, \inm{f.m \at g}{+}{i}}
{\astepc{\comp(i,x)}{f.m \at g}{\comp(i,x')}}
\quad
\Rule
{\astepp{x}{\passive f.m \at g}{x'},\, \inm{g.m \at f}{-}{i}}
{\astepc{\comp(i,x)}{\passive f.m \at g}{\comp(i,x')}}
\quad
\Rule
{\astepp{x}{f.m \uat g}{x'}}
{\astepc{\comp(i,x)}{f.m \uat g}{\comp(i,x')}}
\\
\Rule
{\atermc{u}{f.m \at g},\, \inifp{f.m \at g}{u \parc v}}
{\astepc{u \parc v}{f.m \at g}{v}}
\quad
\Rule
{\atermc{u}{\passive f.m \at g},\, \inifn{g.m \at f}{u \parc v}}
{\astepc{u \parc v}{\passive f.m \at g}{v}}
\quad
\Rule
{\atermc{u}{f.m \uat g}}
{\astepc{u \parc v}{f.m \uat g}{v}}
\\
\Rule
{\atermc{v}{f.m \at g},\, \inifp{f.m \at g}{u \parc v}}
{\astepc{u \parc v}{f.m \at g}{u}}
\quad
\Rule
{\atermc{v}{\passive f.m \at g},\, \inifn{g.m \at f}{u \parc v}}
{\astepc{u \parc v}{\passive f.m \at g}{u}}
\quad
\Rule
{\atermc{v}{f.m \uat g}}
{\astepc{u \parc v}{f.m \uat g}{u}}
\\
\Rule
{\astepc{u}{f.m \at g}{u'},\, \inifp{f.m \at g}{u \parc v}}
{\astepc{u \parc v}{f.m \at g}{u' \parc v}}
\quad
\Rule
{\astepc{u}{\passive f.m \at g}{u'},\, \inifn{g.m \at f}{u \parc v}}
{\astepc{u \parc v}{\passive f.m \at g}{u' \parc v}}
\quad
\Rule
{\astepc{u}{f.m \uat g}{u'}}
{\astepc{u \parc v}{f.m \uat g}{u' \parc v}}
\\
\Rule
{\astepc{v}{f.m \at g}{v'},\, \inifp{f.m \at g}{u \parc v}}
{\astepc{u \parc v}{f.m \at g}{u \parc v'}}
\quad
\Rule
{\astepc{v}{\passive f.m \at g}{v'},\, \inifn{g.m \at f}{u \parc v}}
{\astepc{u \parc v}{\passive f.m \at g}{u \parc v'}}
\quad
\Rule
{\astepc{v}{f.m \uat g}{v'}}
{\astepc{u \parc v}{f.m \uat g}{u \parc v'}}
\\
\RuleC
{\atermc{u}{a},\, \atermc{v}{b}}
{\atermc{u \parc v}{c}}
{a \commm b = c}
\quad
\RuleC
{\atermc{u}{a},\, \astepc{v}{b}{v'}}
{\astepc{u \parc v}{c}{v'}}
{a \commm b = c}
\\
\RuleC
{\astepc{u}{a}{u'},\, \atermc{v}{b}}
{\astepc{u \parc v}{c}{u'}}
{a \commm b = c}
\quad
\RuleC
{\astepc{u}{a}{u'},\, \astepc{v}{b}{v'}}
{\astepc{u \parc v}{c}{u' \parc v'}}
{a \commm b = c}
\\
\Rule
{\atermp{x}{f.m \at g},\, \inm{f.m \at g}{+}{i}}
{\atermp{\iencap{i}(x)}{f.m \at g}}
\quad
\Rule
{\atermp{x}{\passive f.m \at g},\, \inm{g.m \at f}{-}{i}}
{\atermp{\iencap{i}(x)}{\passive f.m \at g}}
\quad
\Rule
{\atermp{x}{f.m \uat g}}
{\atermp{\iencap{i}(x)}{f.m \uat g}}
\\
\Rule
{\astepp{x}{f.m \at g}{x'},\, \inm{f.m \at g}{+}{i}}
{\astepp{\iencap{i}(x)}{f.m \at g}{\iencap{i}(x')}}
\quad
\Rule
{\astepp{x}{\passive f.m \at g}{x'},\, \inm{g.m \at f}{-}{i}}
{\astepp{\iencap{i}(x)}{\passive f.m \at g}{\iencap{i}(x')}}
\quad
\Rule
{\astepp{x}{f.m \uat g}{x'}}
{\astepp{\iencap{i}(x)}{f.m \uat g}{\iencap{i}(x')}}
\end{ruletbl}
\end{table}

The following uniqueness property of the relations
$\inm{f.m \at g\!}{N}{{}}$ will be used in Section~\ref{sect-model} to
construct a model of \ACC+\REC.
\begin{lemma}
\label{lemma-uniqueness}
Let $f,g \in \Loci$ and $m \in \Meth$.
Then for all $I \in \cT_\IF$, there exists an $N \in \cT^\sINT_\Int$
such that for all $N' \in \cT^\sINT_\Int$ with $\inm{f.m \at g}{N'}{I}$
we have that $N = N'$ holds in the initial model of \INT.
\end{lemma}
\begin{proof}
Straightforward, by induction on the structure of $I$.
\qed
\end{proof}

A \emph{bisimulation} $B$ is a triple of symmetric binary relations
$B_\Proc \subseteq \cT_\Proc \x \cT_\Proc$,
$B_\IF \subseteq \cT_\IF \x \cT_\IF$, and
$B_\Comp \subseteq \cT_\Comp \x \cT_\Comp$
such that:
\begin{itemize}
\item
if $B_\Proc(P_1,P_2)$ and $\atermp{P_1}{a}$, then $\atermp{P_2}{a}$;
\item
if $B_\Proc(P_1,P_2)$ and $\astepp{P_1}{a}{P_1'}$, then there exists a
$P_2' \in \cT_\Proc$ such that $\astepp{P_2}{a}{P_2'}$ and
$B_\Proc(P_1',P_2')$;
\item
if $B_\IF(I_1,I_2)$ and $\inm{f.m \at g}{N_1}{I_1}$, then there exists
an $N_2 \in \cT^\sINT_\Int$ such that $\inm{f.m \at g}{N_2}{I_2}$ and
$N_1 = N_2$;
\item
if $B_\Comp(C_1,C_2)$ and $\hasif{C_1}{I_1}$, then there exists an
$I_2 \in \cT_\IF$ such that $\hasif{C_2}{I_2}$ and $B_\IF(I_1,I_2)$;
\item
if $B_\Comp(C_1,C_2)$ and $\atermc{C_1}{a}$, then $\atermc{C_2}{a}$;
\item
if $B_\Comp(C_1,C_2)$ and $\astepc{C_1}{a}{C_1'}$, then there exists a
$C_2' \in \cT_\Comp$ such that $\astepc{C_2}{a}{C_2'}$ and
$B_\Comp(C_1',C_2')$.
\end{itemize}
Let $S \in \set{\Proc,\IF,\Comp}$, and let $t_1,t_2 \in \cT_S$.
Then $t_1$ and $t_2$ are \emph{bisimilar}, written $t_1 \bisim t_2$, if
there exists a bisimulation $B$ such that $B_S(t_1,t_2)$.

The following congruence property of bisimilarity will be used in
Section~\ref{sect-model} to construct a model of \ACC+\REC.
\begin{theorem}[Congruence]
\label{theorem-congruence}
Bisimilarity is a congruence with respect to the operators of
\textup{\ACC+\REC} to build terms of sort $\Proc$, $\IF$ or $\Comp$.
\end{theorem}
\begin{proof}
In the terminology of~\cite{Mid01a}, $\Int$ is a given sort and the
relations ${\inm{f.m \at g\!}{N}{}}$, one for each
$N \in \cT^\sINT_\Int$, constitute a relation parametrized by closed
terms of the sort $\Int$.
Because $\Int$ is a given sort, we can safely identify closed terms of
sort $\Int$ that are semantically equivalent and replace the third
property of bisimulations given above to:
\begin{itemize}
\item
if $B_\IF(I_1,I_2)$ and $\inm{f.m \at g}{N}{I_1}$, then
$\inm{f.m \at g}{N}{I_2}$.
\end{itemize}
Because the relations ${\inm{f.m \at g\!}{N}{}}$ constitute a relation
parametrized by closed terms of a given sort, we can safely replace the
rules for the operational semantics with the conclusions
$\inm{f.m \at g}{+}{i}$ and $\inm{f.m \at g}{-}{i}$ by the rules
\begin{ldispl}
\RuleC
{\inm{f.m \at g}{N}{i}}
{\inm{f.m \at g}{+}{i}}
{\sg(N) = 1}
\quad \mathrm{and} \quad
\RuleC
{\inm{f.m \at g}{N}{i}}
{\inm{f.m \at g}{-}{i}}
{\sg(N) = -1}\;,
\end{ldispl}
where $N$ stands for an arbitrary closed term from $\cT^\sINT_\Int$.
By these replacements, bisimilarity becomes an instance of bisimilarity
by the definition given in~\cite{Mid01a} and the rules for the
operational semantics of \ACC+\REC\ become a complete transition system
specification in panth format by the definitions given in~\cite{Mid01a}.
Hence, it follows by Theorem~4 from~\cite{Mid01a} that bisimilarity is
a congruence with respect to all operators of \ACC+\REC\ to build terms
of sort $\Proc$, $\IF$ or $\Comp$.
\qed
\end{proof}

\section{A Bisimulation Model of \ACC+\REC}
\label{sect-model}

In this section, we construct a model of \ACC+\REC\ using the notion of
bisimilarity defined in Section~\ref{sect-bisim}.
It is a model in which all processes are finitely branching, i.e.\
they have at any stage only finitely many alternatives to proceed.

Henceforth, we will write $\gI_\INT$ for the initial model of \INT, and
$\bbZ$ for the set associated with the sort $\Int$ in $\gI_\INT$.

The \emph{bisimulation model} $\gB_{\ACC+\REC}$ is the expansion of
$\gI_\INT$, the initial model of \INT, with
\begin{itemize}
\item
for each sort $S \in \set{\Proc,\IF,\Comp}$, the set $\cT_S / {\bisim}$;
\item
for each constant $\const{\op_0}{S}$
of \ACC+\REC\ with $S \in \set{\Proc,\IF,\Comp}$, the element
$\opi_0 \in \cT_S / {\bisim}$ defined by $\opi_0 = \beqvc{\op_0}$;
\item
for each operator $\funct{\op_1}{S}{S'}$
of \ACC+\REC\ with $S,S' \in \set{\Proc,\IF,\Comp}$, the operation
$\funct{\opi_1}{\cT_{S} / {\bisim}}{\cT_{S'} / {\bisim}}$ defined by
$\opi_1(\beqvc{t}) = \beqvc{\op_1(t)}$;
\item
for each operator $\funct{\op_2}{S \x S'}{S''}$
of \ACC+\REC\ with $S,S',S'' \in \set{\Proc,\IF,\Comp}$, the operation
$\funct{\opi_2}{\cT_{S} / {\bisim} \x \cT_{S'} / {\bisim}}
       {\cT_{S''} / {\bisim}}$ defined by
$\opi_2(\beqvc{t_1},\beqvc{t_2}) = \beqvc{\op_2(t_1,t_2)}$;
\item
for each operator $\funct{\mult{f.m \at g}}{\IF}{\Int}$
with $f,g \in \Loci$ and $m \in \Meth$, the operation
$\funct{\multi{f.m \at g}}{\cT_{\IF} / {\bisim}}{\bbZ}$ defined by
$\multi{f.m \at g}(\beqvc{I})$ is the unique interpretation in
$\gI_\INT$ of all $N \in \cT^\sINT_\Int$ for which
$\inm{f.m \at g}{N}{I}$.
\end{itemize}

The well-definedness of the operations associated with the operators of
\ACC+\linebreak[2]\REC\ in $\gB_{\ACC+\REC}$ follows immediately from
Theorem~\ref{theorem-congruence}, except for the operations associated
with the operators $\mult{f.m \at g}$.
The well-definedness of the operations associated with the operators
$\mult{f.m \at g}$ in $\gB_{\ACC+\REC}$ follows immediately from
Lemma~\ref{lemma-uniqueness} and the definition of bisimilarity.

We have the following soundness result.
\begin{theorem}[Soundness]
Let $S \in \set{\Int,\Proc,\IF,\Comp}$ and let $t,t' \in \cT_S$.
Then $t = t'$ is derivable from the axioms of \textup{\ACC+\REC} only if
$t = t'$ holds in $\gB_{\ACC+\REC}$.
\end{theorem}
\begin{proof}
It is sufficient to prove the soundness of each axiom separately.
Because $\gB_{\ACC+\REC}$ is an expansion of $\gI_\INT$, it is not
necessary to prove the soundness of the axioms of \INT.
For each of the remaining axioms except M1--M5, soundness is easily
proved by constructing a witnessing bisimulation (for the witnessing
bisimulations for the axioms of \ACP+\REC, see e.g.~\cite{BV95a}).
What remains are the proofs for axioms M1--M5.
The soundness of these axioms follow immediately from the definition of
$\multi{f.m \at g}$ and the rules of the operational semantics.
% concerning the relations $\inm{f.m \at g}{N}{{}}$.
\qed
\end{proof}

\section{Localized Processes}
\label{sect-localized}

If processes are looked at in isolation, it is convenient to abstract
from the loci at which they reside.
This brings us to consider processes made up of actions of the forms
$f.m$ and $\passive f.m$.
These processes are called localized processes.
In this section, we extend \ACC\ with localized processes.
The resulting theory is called \ACClp.

Henceforth, actions from $\Act$ will also be called non-localized
actions, and processes made up of actions from $\Act$ will also be
called non-localized processes.

In \ACClp, we have, in addition to the set $\Act$ of non-localized
actions, the set $\Actl$ of \emph{localized actions} consisting of:
\begin{itemize}
\item
for each $f \in \Loci$ and $m \in \Meth$,
the \emph{active localized action} $f.m$;
\item
for each $f \in \Loci$ and $m \in \Meth$,
the \emph{passive localized action} $\passive f.m$.
\end{itemize}
Intuitively, these localized actions can be explained as follows:
\begin{itemize}
\item
$f.m$ is the action by which a localized process requests a process
residing at locus $f$ to carry out method $m$;
\item
$\passive f.m$ is the action by which a localized process grants a
request of a process residing at locus $f$ to carry out method $m$.
\end{itemize}
It is not possible to perform localized actions synchronously.

Different from \ACC, \ACClp\ has two sorts of processes.
That is, \ACClp\ has the sorts $\Comp$, $\Proc$, $\IF$ and $\Int$ from
\ACC, and in addition the sort $\Procl$ of \emph{localized processes}.
To build terms of sort $\Comp$, \ACClp\ has the constants and operators
of \ACC\ to build terms of sort $\Comp$.
To build terms of sort $\Proc$, \ACClp\ has the constants and operators
of \ACC\ to build terms of sort $\Proc$ and in addition the following
operators:
\begin{itemize}
\item
for each $f \in \Loci$, the unary \emph{placement} operator
$\funct{\atl{f}}{\Procl}{\Proc}$.
\end{itemize}
To build terms of sort $\Procl$, \ACClp\ has the following constants and
operators:
\begin{itemize}
\item
the \emph{deadlock} constant $\const{\dead}{\Procl}$;
\item
for each $a \in \Actl$, the \emph{localized action} constant
$\const{a}{\Procl}$;
\item
the binary \emph{alternative composition} operator
$\funct{\altc}{\Procl \x \Procl}{\Procl}$;
\item
the binary \emph{sequential composition} operator
$\funct{\seqc}{\Procl \x \Procl}{\Procl}$;
\item
the binary \emph{parallel composition} operator
$\funct{\parc}{\Procl \x \Procl}{\Procl}$;
\item
the binary \emph{left merge} operator
$\funct{\leftm}{\Procl \x \Procl}{\Procl}$;
\item
for each $H \subseteq \Act$,
the unary \emph{encapsulation} operator
$\funct{\encap{H}}{\Procl}{\Procl}$.
\end{itemize}
To build terms of sort $\IF$, \ACClp\ has the constants and operators of
\ACC\ to build terms of sort $\IF$.
To build terms of sort $\Int$, \ACClp\ has the constants and operators of
\ACC\ to build terms of sort $\Int$.

Terms of the different sorts are built as usual for a many-sorted
signature.
We assume that there are infinitely many variables of sort $\Procl$,
including $r$, $s$, $r'$ and $s'$.

The constants and operators to build terms of sort $\Procl$ need no
further explanation.
They differ from the constants and operators to build terms of sort
$\Proc$ in that:
(i)~the (non-localized) action constants are replaced by the localized
action constants and
(ii)~the communication merge operator $\commm$ is removed.

Let $L$ be a closed term of sort $\Procl$.
Intuitively, the operators $\atl{f}$ can be explained as follows:
\begin{itemize}
\item
$\atl{f}(L)$ behaves as $L$ with each action $g.m$ replaced by
$g.m \at f$ and each action $\passive g.m$ replaced by
$\passive g.m \at f$.
\end{itemize}
In other words, $\atl{f}$ turns localized processes into non-localized
processes by placing them as a whole in locus $f$.

The axioms of \ACClp\ are the axioms of \ACC, the axioms given in
Tables~\ref{axioms-ACClp1} and~\ref{axioms-ACClp2},%
\begin{table}[!t]
\caption{Axioms for placement of localized processes}
\label{axioms-ACClp1}
\begin{eqntbl}
\begin{axcol}
\atl{f}(\dead) = \dead                             & \axiom{P1} \\
\atl{f}(g.m) = g.m \at f                           & \axiom{P2} \\
\atl{f}(\passive g.m) = \passive g.m \at f         & \axiom{P3} \\
\atl{f}(r \altc s) = \atl{f}(r) \altc \atl{f}(s)   & \axiom{P4} \\
\atl{f}(r \seqc s) = \atl{f}(r) \seqc \atl{f}(s)   & \axiom{P5}
\end{axcol}
\end{eqntbl}
\end{table}
\begin{table}[!t]
\caption{Axiom for parallel composition of localized processes}
\label{axioms-ACClp2}
\begin{eqntbl}
\begin{axcol}
r \parc s = r \leftm s \altc s \leftm r            & \axiom{M1}
\end{axcol}
\end{eqntbl}
\end{table}
and copies of axioms A1--A7, CM2--CM4 and D1--D4 from
Table~\ref{axioms-ACP} with $x$, $y$ and $z$ replaced by different
variables of sort $\Procl$, $a$ standing for an arbitrary constant of
sort $\Procl$ and $H$ standing for an arbitrary subset of $\Actl$.
%b
Axioms P1--P5 are the defining axioms of $\atl{f}$.
Axiom M1 replaces axiom CM1.
The latter axiom is not suited for the localized case because it is not
possible to perform localized actions synchronously.

Guarded recursion can be added to \ACClp\ as it is added to \ACP\ in
Section~\ref{sect-REC}.
We write \ACClp+\REC\ for \ACClp\ extended with the constants standing
for the unique solutions of guarded recursive specifications and the
axioms RDP and RSP.

As an example of a localized process, we give the localized buffer
process $B'$ defined by the guarded recursion equation
\begin{ldispl}
B' = \Altcv{d \in \cD} \passive f.c_d \seqc f.c_d \seqc B'\;.
\end{ldispl}
If $g$ and $h$ are different loci, then the processes $\atl{g}(B')$ and
$\atl{h}(B')$ reside at different loci, but apart from that they are the
same.
The connection between $B'$ and the buffer process $B$ defined in
Section~\ref{sect-another-example} is couched in the equation
$B = \atl{g}(B')$, which is derivable from the axioms of \ACClp+\REC.
The placement operators are primarily useful in cases where `copies' of
the same process coexist at different loci.
However, they are also useful otherwise to obtain more terse
descriptions of processes.
Much more complicated processes than buffers with capacity one are
needed to illustrate this.

In the structural operational semantics of \ACClp+\REC, the following
relations are used in addition to the ones used in the structural
operational semantics of \ACC+\REC:
\begin{itemize}
\item
a unary relation ${\termlp{a}} \subseteq \cT_\Procl$,
for each $a \in \Actl$;
\item
a binary relation
${\steplp{a}} \subseteq \cT_\Procl \x \cT_\Procl$,
for each $a \in \Actl$.
\end{itemize}
We write
$\atermlp{L}{a}$ instead of $L \in {\termlp{a}}$ and
$\asteplp{L}{a}{L'}$ instead of $\tup{L,L'} \in {\steplp{a}}$.
The relations can be explained as follows:
\begin{itemize}
\item
$\atermlp{L}{a}$:
localized process $L$ is capable of first performing $a$ and then
terminating successfully;
\item
$\astepp{L}{a}{L'}$:
localized process $P$ is capable of first performing $a$ and then
proceeding as localized process $P'$.
\end{itemize}

The structural operational semantics of \ACClp+\REC\ is described by the
rules for the operational semantics of \ACC+\REC, the rules given in
Table~\ref{rules-ACClp}, and copies of the rules without the
side-condition $a \commm b = c$ from Table~\ref{rules-ACP+REC} with
${\termp{a}}$ and ${\stepp{a}}$ replaced by ${\termlp{a}}$ and
${\steplp{a}}$, respectively, $x$, $x'$, $y$ and $y'$ replaced by
different variables of sort $\Procl$, $a$ standing for an arbitrary
constant of sort $\Procl$ and $H$ standing for an arbitrary subset of
$\Actl$.
\begin{table}[!t]
\caption{Additional rules for operational semantics of \ACClp}
\label{rules-ACClp}
\begin{ruletbl}
\Rule
{\atermlp{r}{g.m}}
{\atermp{\atl{f}(r)}{g.m \at f}}
\quad
\Rule
{\atermlp{r}{\passive g.m}}
{\atermp{\atl{f}(r)}{\passive g.m \at f}}
\quad
\Rule
{\asteplp{r}{g.m}{r'}}
{\astepp{\atl{f}(r)}{g.m \at f}{\atl{f}(r')}}
\quad
\Rule
{\asteplp{r}{\passive g.m}{r'}}
{\astepp{\atl{f}(r)}{\passive g.m \at f}{\atl{f}(r')}}
\end{ruletbl}
\end{table}

Constructing a bisimulation model of \ACClp+\REC\ can be done on
the same lines as constructing a bisimulation model of \ACC+\REC.

\section{Conclusions}
\label{sect-conclusions}

In this paper, we have built on earlier work on \ACP\ and earlier work
on interface groups.
\ACP\ was first presented in~\cite{BK84b} and interface groups were
proposed in~\cite{BP07a}.
We have introduced an interface group for process components and have
presented a theory about process components of which that interface
group forms part.
The presented theory is a development on top of \ACP.
We have illustrated the use of the theory by means of examples, and
have given a bisimulation semantics for process components which
justifies the axioms of the theory.

Two interesting properties of the interface group for process components
introduced in this paper are:
(i)~the interface combination operator $+$ is not idempotent and
(ii)~for each $f,g \in \Loci$ and $m \in \Meth$, the interface element
constants $f.m \at g$ and $\passive g.m \at f$ are each other inverses.
Property~(i) allows for expressing that a process component expects from
a number of process components an ability or promises a number of
process components an ability.
Property~(ii) allows for establishing on the basis of its interface that
a process component composed of other process components is a closed
system.

Like  in~\cite{BP07a}, the inclusion of behavioural information in
component interfaces has been deliberately rejected in order to have
orthogonality between component interfaces and component behaviours.
The distinction between active interface elements and passive interface
elements made in this paper corresponds to the distinction between
import services and export services made in~\cite{Pah03a}.
Adaptations of module algebra~\cite{BHK90a} that allow for this kind of
distinction are investigated in~\cite{FQ02a}.
However, interface groups are not considered in those investigations.

Processes as considered in \ACP\ have been combined with interfaces
before in $\mu$CRL~\cite{GP95a} and PSF~\cite{MV90a}, two tool-supported
formalisms for the description and analysis of processes with data.
However, in $\mu$CRL and PSF, interfaces serve for determining whether
descriptions of processes are well-formed only.

\bibliographystyle{spmpsci}
\bibliography{PA}

% \par \vfill \par \noindent DRAFT of \today

\end{document}